\documentclass[twocolumn,twoside,slac_two]{revtex4}
\usepackage{graphicx}
\usepackage{fancyhdr}
\begin{document}

\title{Short time scale variability at gamma rays in FSRQs and implications on the current models}

%

\author{L. Foschini, G. Ghisellini, F. Tavecchio, G. Bonnoli}
\affiliation{INAF Osservatorio Astronomico di Brera, 23807 Merate, Italy}
\author{A. Stamerra}
\affiliation{Universit\`a di Siena, 53100 Siena, Italy}

\begin{abstract}
We studied the rapid variability at GeV gamma rays of the flat-spectrum radio quasar PKS~1222$+$216, which was recently found by the MAGIC Cerenkov telescope to display very short variability (minutes time scale) at hundreds of GeV. We analyzed the time period between 2010 April 29 and June 20, when the source generated a few $\gamma$-ray flares with flux in the MeV-GeV band in excess of $10^{-5}$~ph~cm$^{-2}$~s$^{-1}$ on daily basis. We set tight upper limits on the observed doubling time scale ($\sim 1$~hour on 2010 April 30), the smallest measured to date at MeV-GeV energies, which can constrain the size of the $\gamma$-ray emitting region. We also studied the spectra measured during two flares (2010 April 30 and June 17-18). The combination of spectral and variability studies obtained in the present work favors the hypothesis that $\gamma$ rays are generally produced inside the broad-line region (BLR), but sometimes the dissipation can occur at larger distances, nearby the infrared torus.
\end{abstract}

\maketitle

\thispagestyle{fancy}


\section{INTRODUCTION}
The variability of the emission from active galactic nuclei (AGN) with relativistic jets viewed at small angles is amplified because of the Doppler boosting $\delta$. This measurement can be linked to the size of the emitting region $r$, which should be smaller than $\tau \delta c/(1+z)$, being $\tau$ the observed characteristic time scale, $c$ the speed of light, and $z$ the redshift of the source. Rather obviously, very short time scales imply very small emitting regions. When the size $r$ is smaller than the gravitational radius of the central black hole ($r_{\rm g}=GM/c^2$, where $G$ is the gravitational constant and $M$ is the mass of the singularity), this can create some problems in blazar models (cf \cite{BEGELMAN}). According to the present models, the dissipation region of jets in blazars is at about $10^{3}r_{\rm g}$ (e.g. \cite{GHISELLINI}). By taking into account that any perturbation propagate with a starting size of the order of the gravitational radius and the jet is a self-similar structure -- which in turn implies that the emitting region is linked to the distance from the central black hole $R$ through the aperture of the jet $\psi\sim 0.1-0.25$ (e.g. \cite{DERMER,MODEL}) -- the dissipation at $R\sim 10^{3}r_{\rm g}$ requires an emitting region of the order of $(100-250)r_{\rm g}$. Such a size can sustain a variability of about $7-17$~hours in the case of a blazar at $z=1$, $M=10^{9}M_{\odot}$, and $\delta=10$. 

The observation of sub-hour time scales in the high-energy emission from some blazars (\cite{FOSCHINI0,HESS,MAGIC1,MAGIC2}) threatened this scenario. Although a simple increase of the Doppler factor could be an affordable gross solution in the case of BL Lac Objects (which have a photon-starved nearby environment) this is no more reasonable for flat-spectrum radio quasars (FSRQs), with a richer photon field. Since the external energy density is proportional to $\Gamma^2$ ($\Gamma$ is the bulk Lorentz factor), an increase of $\delta$ implies an increase of the energy density and, hence, of the optical depth for pair production (cf \cite{MODEL}). Particularly, the observation of variability with time scales of minutes at hundreds of GeV in the case of PKS~1222$+$216 \cite{MAGIC2} has put very severe constraints to the existing models (see \cite{TAVECCHIO} for a review and some possible solutions). 

Therefore, the search for short time scales in the emission of FSRQs is an insuperable testing ground for the existing theories about relativistic jets. The launch of the Large Area Telescope (LAT, \cite{LAT}) onboard the {\it Fermi} satellite made it available the state-of-the-art of the high-energy $\gamma$-ray instrumentation. With its superior performance, it is now possible to probe the variability in FSRQs on hourly time scale. Some work has been already published in \cite{FOSCHINI2,FOSCHINI1}. Here we focus on the case of PKS~1222$+$216 as observed by {\it Fermi}/LAT and reanalyzed the data presented in \cite{FOSCHINI1} with a most recent and improved version of the software.

\section{DATA ANALYSIS}
The FSRQ PKS~1222$+$216 (a.k.a. 4C~$+$21.35) is a high-power blazar at $z=0.432$. It was listed in the catalog of very high energy ($E>100$~GeV) sources detected by {\it Fermi}/LAT (in a time period of about two years) \cite{NERONOV}. In 2010 April-June, the source underwent a few strong $\gamma$-ray outbursts, with fluxes at energies greater than 100~MeV in excess of $10^{-5}$~ph~cm$^{-2}$~s$^{-1}$ \cite{TANAKA}. Particularly, on 2010 June 17, it was detected at energies of hundreds of GeV by the MAGIC Cerenkov telescope, while displaying very short time scale variability ($8.6_{-0.9}^{+1.1}$~minutes, \cite{MAGIC2}). 

The analysis of {\it Fermi}/LAT data in the period 2010 April-June revealed short variability: $1.5\pm0.6$~hours or $<2.3$~hours, with a more conservative approach \cite{FOSCHINI1}. This analysis was performed by using a pre-flight version of the instrument response function (IRF), adapted to overcome some problems in the $100-200$~MeV energy range. After the publication of the work \cite{FOSCHINI1}, the {\it Fermi}/LAT Collaboration released (2011 August 5) a new fully post-flight IRF, together with improved files for the background subtraction. Given the importance of this release, we decided to reanalyze the LAT data, searching for tighter estimates of the time variability. 

\begin{figure}[!t]
\centering
\includegraphics[angle=270,scale=0.33]{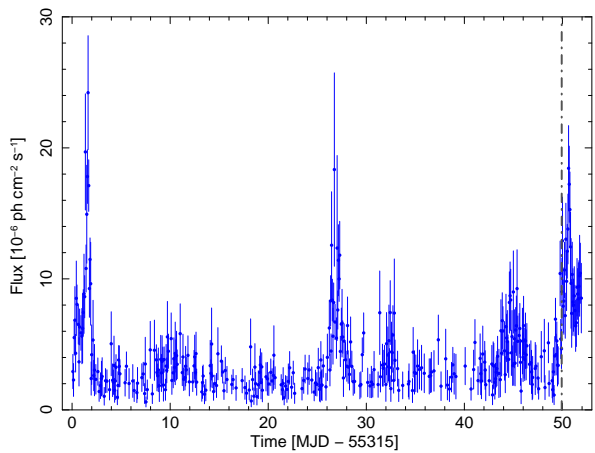}
\caption{{\it Fermi}/LAT light curve in the $0.1-100$~GeV energy band with GTI time bins (too small to be visible). The vertical grey dot-dashed line indicates the period of the MAGIC detection \cite{MAGIC2} (see the zoom in Fig.~\ref{fig:zoom2}). Time starts on MJD 55315 (2010 April 29). } 
\label{fig:total}
\end{figure}

Therefore, in this work, we adopt the LAT {\tt Science Tools v. 9.23.1}, with the IRF {\tt P7SOURCE\_V6}, and background files {\tt iso\_p7v6source.txt} (isotropic) and {\tt gal\_2yearp7v6\_v0.fits} (Galactic diffuse). We selected events of class 2, within a zenithal distance of $100^{\circ}$ and a rocking angle smaller than $52^{\circ}$. The time period covered by the analysis is the same of \cite{FOSCHINI1}, i.e. 2010 April 29 (MJD 55315) and June 20. The remaining of the analysis has been done as described in \cite{FOSCHINI1}, i.e. by building a light curve with the width of bins equal to the Good-Time-Intervals (GTI). Then, we searched for the time scale $\tau$ for doubling/halving the flux as defined by:

\begin{equation}
\frac{F(t)}{F(t_0)}=2^{\frac{-(t-t_0)}{\tau}}
\label{equation}
\end{equation}

\noindent where $F(t)$ and $F(t_0)$ are the fluxes at times $t$ and $t_0$, respectively. Fig.~\ref{fig:total} shows the $\gamma$-ray light curve of PKS~1222$+$216 in the analyzed period. 

The analysis based on the calculation of $\tau$ from two consecutive points and Eq.~(\ref{equation}) results only in an upper limit of $\tau <2.8$~hours, similar to the $\tau <2.3$~hours obtained with the previous analysis. 

\begin{figure}[!t]
\centering
\includegraphics[angle=270,scale=0.33]{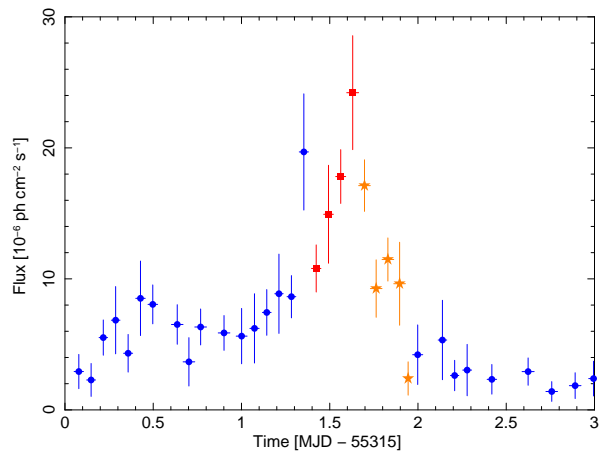}
\caption{Zoom on the first flare occurred on 2010 April 30 (MJD 55316). Time starts on MJD 55315 (2010 April 29). See the text for details.} 
\label{fig:zoom1}
\end{figure}

\begin{figure}[!t]
\centering
\includegraphics[angle=270,scale=0.33]{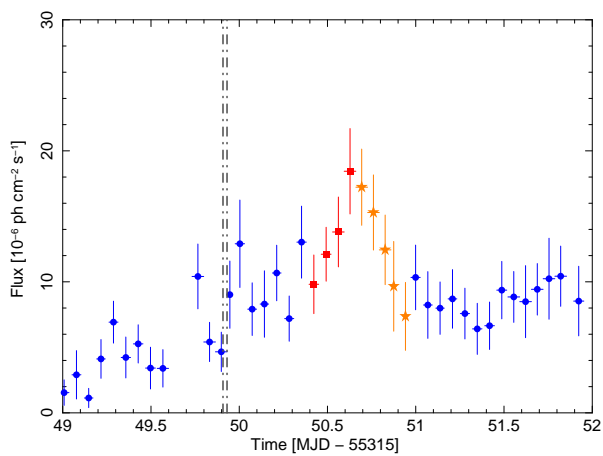}
\caption{Zoom on the period of the MAGIC detection occurred on 2010 June 17 (MJD 55364.908, indicated with grey dot-dashed vertical lines \cite{MAGIC2}). It is worth noting that {\it Fermi} detected the highest flux {\it after} the MAGIC observation.  Time starts on MJD 55315 (2010 April 29). See the text for details.} 
\label{fig:zoom2}
\end{figure}

However, in some cases like the flares occurred on 2010 April 30 and June 17-18 (see Fig.~\ref{fig:zoom1} and \ref{fig:zoom2}), it is possible to perform a fit on a few consecutive points, thus improving the measurement errors. The best estimates can be obtained during the flare of 2010 April 30 15:07 UTC (MJD 55316.6301, Fig.~\ref{fig:zoom1}), which reached a peak flux of $(2.4\pm 0.4)\times 10^{-5}$~ph~cm$^{-2}$~s$^{-1}$. The rise time is calculated with four points between MJD 55136.42 and 55136.63 (i.e. the points betwee 1.42 and 1.63 in the abscissa of Fig.~\ref{fig:zoom1} indicated with the red squares) and resulted in $\tau = 4.5\pm 1.6$~hours. The decay is sharper and, with six points between 1.63 and 1.94 (i.e. MJD 55136.63 and 55136.94; see the orange stars plus the peak in Fig.~\ref{fig:zoom1}), it is possible to calculate a $\tau = -1.0\pm 0.2$~hours. 

Another $\gamma$-ray flare occurred {\it after} the MAGIC observation \cite{MAGIC2} (Fig.~\ref{fig:zoom2}), with a peak of $(1.8\pm 0.3)\times 10^{-5}$~ph~cm$^{-2}$~s$^{-1}$ on MJD 55365.6276 (2010 June 18 15:03 UTC). The rise time derived from the fit of four points (red squares in Fig.~\ref{fig:zoom2}) is $\tau = 6.1\pm 3.1$~hours, while the decay is $\tau = -4.7\pm 1.5$~hours (fit with six points, orange stars plus the peak in Fig.~\ref{fig:zoom2}). 

\begin{figure}[!t]
\centering
\includegraphics[angle=270,scale=0.33]{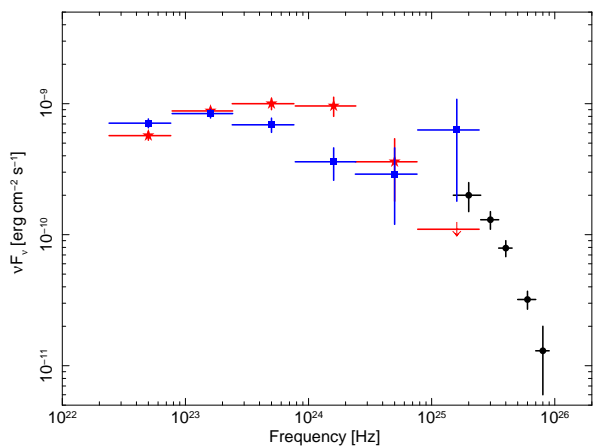}
\caption{$\gamma$-ray spectral energy distribution (SED) of PKS~1222$+$216. The MAGIC detection of 2010 June 17 is indicated with black circles (from \cite{MAGIC2}). {\it Fermi}/LAT detection of 2010 April 30 (1 day integration) is displayed with red stars (the upper limit is at $5\sigma$ level), while blue triangles are for the detection on 2010 June 18 (1 day integration).} 
\label{fig:sed}
\end{figure}

It is worth noting that the MAGIC observation refers to 2010 June 17 \cite{MAGIC2}. The two events (April 30 and June 17-18) are not necessarily correlated or similar. It is not possible to know if on April 30 the source extended its emission to hundreds of GeV, because MAGIC started observing PKS~1222$+$216 on May 3 \cite{MAGIC2}. However, the spectral analysis of {\it Fermi}/LAT data suggests that the emission drops after a few tens of GeV (Fig.~\ref{fig:sed}, red star points). The photon with highest energy detected by LAT at that time measured $\sim 23$~GeV. Instead, the spectrum measured during the flare of 2010 June 17-18 (blue squares in Fig.~\ref{fig:sed}) seems to extend to match the MAGIC measurements of \cite{MAGIC2}, with the highest energy of LAT photons of $\sim 63$~GeV. Particularly, in the last {\it Fermi} bin ($31<E<100$~GeV), the source is detected with high significance ($TS = 30$, equivalent to $\sim 5.5\sigma$). The partial drop in flux as observed in the two $3<E<10$~GeV and $10<E<31$~GeV bins  can be consistent with an absorption within the BLR according to the theory of Poutanen \& Stern \cite{POUTANEN}, as already noted in \cite{TANAKA}. 

\section{FINAL REMARKS}
From the variability alone, the two flares of 2010 April 30 and June 17-18 seem similar, with the time scales consistent each other within the errors. The value of $\tau \sim 1$~hour favors the dissipation in the BLR (cf. Sect.~3 and Eqs.~3 and 4 of \cite{FOSCHINI1}). However, the spectral analysis revealed two different states of the source: one (2010 April 30) with a spectrum peaking around $\sim 10^{24}$~Hz and with no detection above a few tens of GeV; the other (2010 June 17-18) has a harder spectrum extending to hundreds of GeV, although partially absorbed in the range $\sim 3-31$~GeV. It seems that most of the dissipation generally occurs in the BLR, but sometimes there are episodes where the region is at larger distances from the central spacetime singularity, nearby the infrared torus (see \cite{MAGIC2,TANAKA,TAVECCHIO}). The possibility of an outward motion of the dissipation zone has been already suggested by Stern \& Poutanen \cite{STERN} by studying 3C~454.3. In this case, the motion would be on a larger scale, from the BLR to the infrared torus. 

Perhaps, this is the solution of the long-standing question on where $\gamma$-rays are produced, which has seen several researchers debating during the past years mainly on two opposite arrays: on one side, the two groups led by Marscher \cite{MARSCHER1,MARSCHER2} and Sikora \cite{SIKORA}, respectively, favoring the hypothesis of a location nearby the infrared torus, i.e. on super-pc scales; on the other side, a more heterogeneous group supporting the sub-pc location, within the BLR \cite{FINKE,GHISELLINI,POUTANEN,TAVECCHIO2}. The combined spectral and variability analysis presented in this work suggests that it is not a matter of {\it either} one {\it or} the other hypothesis ({\it aut-aut}), but that {\it both} hypotheses are possible in the same source, depending on the time of the event. That is, the dissipation occurs {\it both} in the BLR {\it and} in the infrared torus. Generally in the BLR, but sometimes the blob is only partially absorbed by the BLR and can move outward still sufficiently collimated and dissipate nearby the torus. 

This is just one case and more coordinated observations in the GeV-TeV range are necessary to understand if this is an isolated anomaly or a typical behavior of blazars. 

\bigskip
\begin{acknowledgements}
{\it Added on February 13, 2012:} The recent preprint by Nalewajko et al. \cite{NALEWAJKO} casted some doubts on our measurements of the characteristic time scales for doubling/halving the fluxes. Indeed, we discovered an error, which led to an underestimate of $\tau$. We have now corrected the error and we thank Nalewajko et al. for having pointed it out. It is worth noting that this error does not affect the concepts expressed in the present work.  
\end{acknowledgements}

\end{document}